\documentstyle[multicol,aps,epsfig]{revtex} 

\tighten       

\begin{document}
\draft

\title{STM induced hydrogen desorption via a hole resonance}
\author{K. Stokbro$^1$, C. Thirstrup$^2$, M. Sakurai$^2$,
  U. Quaade$^1$, Ben Yu-Kuang Hu$^{1}$, F. Perez-Murano$^{1,3}$ and F. Grey$^1$ }
\address{$^1$Mikroelektronik Centret, Danmarks Tekniske Universitet, 
Bygning 345\o , DK-2800 Lyngby, Denmark.}
\address{$^2$Surface and Interface Laboratory, RIKEN, Saitama 351,
Japan.}

\address{$^3$Department of Electronic Engineering Universitat Autonoma de Barcelona,
08193 Bellaterra, Spain.}
\date{\today}
\maketitle

\begin{abstract}
We report STM-induced desorption of H from Si(100)-H(2$\times1$) at
negative sample bias. The desorption rate exhibits a power-law
dependence on current and a maximum desorption rate at
$-7$~V.  The desorption  is explained by  vibrational heating of
H due to inelastic scattering of tunneling holes with
the Si-H~5$\sigma$ hole resonance. The dependence of desorption rate
on current and bias is analyzed using a 
 novel approach for calculating  inelastic
scattering, which includes the effect of the electric field between tip and sample.
We show that the  maximum desorption rate at $-7$~V is due to a maximum
fraction of inelastically scattered electrons  at the onset of the field emission
regime. 
\end{abstract}
\begin{multicols}{2}

\narrowtext
 Single atom
manipulation with a scanning tunneling microscope (STM) has been reported for several  
systems and a variety of physical mechanisms has been proposed to
account for such manipulation\cite{StEi91,Av95}.
Among the works relevant for the present Letter
we mention  single hydrogen atom desorption from the
Si(100)-H(2$\times1$) surface\cite{ShWaAbTuLyAvWa95}, and dissociation of single
O$_2$ molecules on Pt(111)\cite{StReHoGaPeLu97}. These manipulations were performed at
positive sample bias, and the underlying microscopic mechanism has
been related to vibrational heating by inelastic scattering of  tunneling electrons with
an electron  resonance on the sample. There have been theoretical predictions that a
related mechanism may operate at negative sample bias\cite{SaPePa94}, 
which  involve
inelastic scattering of a tunneling hole with  a hole resonance on
the sample. However, to our knowledge there has
been no experimental confirmation of such a mechanism, probably
because high tunnel currents and sample biases are needed to obtain
high inelastic scattering rates with 
low-lying hole resonances (see Fig.~1).  

In this Letter, we present evidence for  a desorption mechanism
involving a hole  resonance, for  STM-induced
hydrogen desorption from the monohydride Si(100)-H(2$\times1$) surface in
ultra-high vacuum\cite{ShWaAbTuLyAvWa95}. Whereas in previous
studies\cite{ShWaAbTuLyAvWa95,AdMaSw96,LyShHuTuAb94}, hydrogen
desorption has been studied  at positive sample bias, here we
report hydrogen desorption at negative bias.  The desorption process is modelled  by  vibrational
heating of hydrogen caused by inelastic scattering of tunneling
holes with
the Si-H $5\sigma$ hole resonance. The inelastic
scattering rates are calculated using a novel method based on first
principles electronic structure theory, and desorption rates obtained
by solving the Pauli master equation for a truncated harmonic
potential well\cite{GaPeLu92}. With this model we obtain desorption
rates as function of tunnel current and sample bias that are in
 agreement with the experimental 
data. We find a maximum desorption rate at sample bias $-7$~V, which 
 coincides with the onset of the field emission or
Fowler-Nordheim regime\cite{Si63,FoNo28}. We show that this  results from a competition
between polarization of the
hole resonance, which  increases the fraction of inelastic scattered
electrons, and domination of Fermi-level contributions to the
tunnel-current  in the field emission regime.

The experiments were performed on n-type  Si(100) ($N_D=1\times
10^{18}\; {\rm cm}^{-3}$) samples using a JEOL JSTM-4000XV microscope at a base pressure
of $1\times 10^{-10}$~torr. 
Atomic hydrogen was absorbed on the clean Si(100)-(2$\times$1)
reconstructed surface to obtain the monohydride (2$\times$1) phase, in a manner
identical to previous reports\cite{LyShHuTuAb94}, and  W tips were
used.  The desorption experiments
were carried out by scanning the STM tip  at  speed, $s$, sample bias, $V_{\rm b}$, and tunnel
current, $I$, and subsequently imaging the affected area to determine
the number of Si sites where desorption occurred. Figure~2 shows a typical
example of desorption at negative sample bias of hydrogen along a
single dimer row. We first measured the dependence of the desorption
rate, $R$,  on the  tunnel current for sample biases of $V_b=-7$~V and
$-5$~V. The results are shown in Fig.~3(inset). For both biases
there is a power-law dependence of the desorption rate upon current,
$R=R_0 (I/I_{\rm des}) ^{\alpha}$, with exponent $\alpha \approx
6$. In this equation $I_{\rm des}$ is the tunnel current that gives
rise to a fixed desorption rate $R_0$. 
In order to find the voltage dependence of the desorption rate, we  measure $I_{\rm des}(V_{\rm b})$ 
with  $V_{\rm b}$ in the range $-10$~V to $-4$~V, as shown in Fig.~3.
 The measurements were obtained by scanning 30~nm along  a
 single dimer row 
at $s=2$~nm/s, and  $I_{\rm des}$ is defined as  the current
that gives rise to desorption of  50\% of the
hydrogen along such a scan line, which corresponds to $R_0=4$~s$^{-1}$.

The main feature of Fig.~3 is that
significantly higher bias and tunnel currents are required at negative
bias, compared to the positive bias case\cite{ShWaAbTuLyAvWa95}. Whereas $I_{\rm des}(V_{\rm b})$ decreases
monotonically at positive $V_{\rm b}$\cite{ShWaAbTuLyAvWa95}, it displays a minimum at negative $V_{\rm b}$,
increasing for $V_{\rm b}<-7$~V. Experimentally, the lithographic resolution at
negative bias is comparable to that at positive bias. At the highest
current levels there is a tendency for the tip resolution to be affected by
the lithography process. This limits the voltage range over which
$I_{\rm des}(V_{\rm b})$ can be readily measured, and may explain why no
detailed investigation of desorption at negative sample bias has been
made previously. We note that the behaviour of $I_{\rm des}(V_{\rm b})$ at negative
sample bias has been confirmed independently on a separate STM system\cite{Peetal97}.

At negative sample bias, electrons accelerated across the tunnel gap
impinge on the tip, so the possibility of direct excitation by
collision can be ruled out as a desorption mechanism. In contrast, at
positive sample bias, direct excitation of the Si-H bond is believed to play the
dominant role for $V_{\rm b}>4$~V\cite{ShWaAbTuLyAvWa95,AdMaSw96}.
Shen {\it et~al.}\cite{ShWaAbTuLyAvWa95} have proposed vibrational
heating of hydrogen\cite{WaNeAv93} by tunneling electrons 
scattering inelastically with the Si-H~6$\sigma^*$  electron resonance as a
desorption mechanism for positive sample 
bias and $V_{\rm b}<4$~V. A similar mechanism can function at negative sample
bias, since an electron  tunneling from the sample to the tip may
excite the Si-H $5\sigma$  hole resonance, which upon deexcitation
can  transfer energy to the hydrogen atom\cite{SaPePa94}. 
This process can be viewed
as  inelastic scattering of a tunneling hole  with the  Si-H $5\sigma$ hole
 resonance, as illustrated in Fig.~1. A characteristic
 feature of vibrational heating by inelastic scattering
 is a power-law dependence of the desorption rate on
 current\cite{GaPeLu97} in agreement with the experimental observations.

To make a quantitative analysis of  the dependence of the desorption rate on tunnel current and
 sample bias, we
 develop a  method for obtaining inelastic
scattering rates from first principles electronic structure calculations. Since the inelastic scattering rate depends on the
tip-sample distance, $d$, while the measured quantities are $I$ and
$V_{\rm b}$, we first calculate the relation between $I$,
$V_{\rm b}$, and $d$. For this purpose we use  a high voltage extension of
 the work  by Tersoff and Hamann\cite{TeHa85}, that includes the
 effect of the electric field between tip and sample\cite{StQuGr97}. 
The tunnel current is obtained from the local density of states(DOS)  of the
Si(100)-H(2$\times1$) surface, $\rho({\bf r},\varepsilon,E)=\sum_{\mu} |\psi_{\mu}({\bf r},E)|^2
\delta(\varepsilon-\varepsilon_{\mu})$, at tip position  ${\bf r}$, and 
 wave functions
$\psi_{\mu}$  for electrons with  energy $\varepsilon_{\mu}$ are
 calculated in an  external field $E$. The electric field 
is modelled by a planar electric field outside the surface and its
magnitude determined from  $V_{\rm b}$ and $d$.  The tunnel current
 is  given by\cite{StQuGr97}
\begin{equation}
 I = C_{\mathrm{w}} \int_{0}^{e V_{\rm b}}
|e^{2R_{\mathrm{w}}\kappa(\varepsilon)}|\rho(d+R_{\mathrm{w}},\varepsilon,E) d\varepsilon,
\end{equation}
where distances are in bohrs, current in Amperes and all electron
energies are  relative to the sample Fermi energy, $\varepsilon_{\rm F}$.
In this equation $\kappa(\varepsilon) = \sqrt{2 m (\phi_t +eV_{\rm b} -\varepsilon)}/\hbar$ is the
wave function inverse decay length, $\phi_t=4.5$~eV the work function of the
W tip\cite{CRC94}, $R_{\mathrm{w}}=3$ a.u. the atomic radius of W, and the
normalization constant $C_{\mathrm{w}}=0.007 R_{\mathrm{w}}^2$ Amperes
$\times$ bohr is obtained from a
calculation of a model W tip\cite{StQuGr97}.

We  test  the method by  calculating  the STM corrugation across a single missing H
defect. The electronic
structure calculations are based on  density-functional theory\cite{HoKo64,KoSh65}  within
the generalized-gradient approximation(GGA)\cite{PeWa91} using 20~Ry plane-wave
basis sets.  The  Si(100)-H(2$\times1$) surface  is represented by a
12 layer ($2\times1$) slab, and we use a  c($4\times4$) slab to 
calculate the corrugation of  a single missing
H defect. Ultra-soft pseudo-potentials\cite{Va90}
are used for both H and Si. Figure~2(b) shows the result compared with
the measured corrugation perpendicular to a 
desorption line. The theoretical corrugation  compares
well with the experimental data when the electric field between tip
and sample is included (dashed line) but not otherwise (dotted line).

We next calculate the tunnel current of inelastically scattered
electrons.   Electrons may scatter inelastically
due to long-range electrostatic interactions  with a vibrational
transition dipole moment, or due to local inelastic scattering with 
an electronic resonance\cite{Pe88}.  Only a local interaction
 can cause   atomic scale desorption, so in the following we
 restrict the analysis to resonance coupling.
The resonance scattering event can be described by a local
polaron model with a linear electron-phonon coupling  $\lambda$, and
  we use a harmonic
approximation for the Si-H bond potential
 with frequency $\omega_0 $. To calculate the inelastic  current,
 $I_n$, which causes 
transitions from vibrational level $0$ to $n$, we  combine
 previous models of inelastic-scattering\cite{SaPePa94,Ga97} with
 the Tersoff-Hamann model of STM tunneling between a surface and a W tip\cite{TeHa85,StQuGr97}. 
 In the limit of a broad resonance, $\Delta \gg \lambda$,$\omega_0$,
 we obtain\cite{Stetal98} 
\begin{equation}
 I_{n} =  C_{\mathrm{w}} \; n! \;\int_{n \hbar\omega_0}^{e V_{\rm b}}
|e^{2 R_{\mathrm{w}} \kappa(\varepsilon) }|\rho_{n}({\bf d}+R_{\mathrm{w}},\varepsilon,E) d\varepsilon ,
\label{eq:iinelas}
\end{equation}
where distances are in a.u. and  $I_{n}$ in Amperes. We define a
dimensionless parameter, 
$K=\pi \lambda^2
\rho_s/\Delta$, where  $\rho_s=\sum_\mu
\delta(\varepsilon_{5\sigma}-\varepsilon_\mu)$  is the average  DOS around the resonance energy
$\varepsilon_{5\sigma}$, and $\Delta$ is  the resonance width. The
weighted local DOS,
\[ 
\rho_n({\bf r},\varepsilon,E) = K^n \sum_{\mu} f_{\mu} |\langle 5\sigma|\mu\rangle |^{2n} |\psi_{\mu}({\bf r},E)|^2
\delta(\varepsilon-\varepsilon_{\mu}),
\] is  weighted with,
$(K|\langle 5\sigma|\mu\rangle|^2)^n$, where $|5\sigma\rangle$ and
$|\mu\rangle$ are the resonance state and sample eigenstates,
respectively, and
 $f_{\mu}=|\langle 5\sigma|\mu\rangle |^{2}/(x+|\langle 5\sigma|\mu\rangle |^{2}) $ is the fraction of
electrons which tunnel from the tip  to state $\mu$ via the $5\sigma$
resonance. The parameter $x$ determines the fraction of electrons
which does not tunnel via the resonance state, and
we have estimated $x\approx 0.1$ $(=0.25 \times
 \mathrm{max}_{\mu}|\langle 5\sigma|\mu\rangle |^{2})$.  We
 note that the calculations are quite insensitive to the
 value of $x$, since   $f_{\mu}\sim
1$ for $\varepsilon_{\mu} \sim \varepsilon_{5\sigma}$ and $f_{\mu}
 \sim |\langle 5\sigma|\mu\rangle |^{2}/x$ for
 $|\varepsilon_{\mu}-\varepsilon_{5\sigma}|\gg \Delta$, thus
 $x$ merely damps  contributions from eigenstates not in
 resonance.

We next calculate the parameters entering Eq.~(2) for 
 inelastic scattering with the Si-H 5$\sigma$  resonance.
 From a frozen phonon calculation we obtain $\hbar \omega_0=0.26$~eV.
Using   the 
ground state energy of a free H atom we calculate $E_{\rm
  des}=3.36$~eV. To find $\langle 5\sigma|\mu\rangle$ 
we  project  the electronic eigenstates of the slab calculation onto the 
5$\sigma$ wave function  of a  Si-H molecule, and the solid line
 in Fig.~4 shows the partial DOS $n_{5\sigma}(\varepsilon)=\sum_{\mu}|\langle 5\sigma|\mu\rangle |^{2}\delta(\varepsilon-\varepsilon_\mu)$. We find that the 
5$\sigma$ resonance  is centered at $\varepsilon_{5\sigma}=-5.3$~eV
 relative to the Fermi level of an $n$-type sample with 
 an average width of
 $\Delta= 0.6$~eV. Crosses in Fig.~4 show $\rho_s(\varepsilon_\mu
 )\approx n(\varepsilon_\mu )/|\langle 5\sigma|\mu\rangle |^{2}$ from
 which we  estimate $\rho_s \approx 1.2 $~eV$^{-1}$. The electron-phonon coupling is given by 
 $\lambda=\varepsilon_{5\sigma}' \sqrt{\hbar/2M\omega_0}$,
where M is the hydrogen mass and $\varepsilon_{5\sigma}'$  the derivative of $\varepsilon_{5\sigma}$ with respect to 
the  Si-H bond length $z$\cite{SaPePa94}. Our calculations show that   
$\varepsilon_{5\sigma}$ varies almost linearly with $z$ in the range
 1.25~\AA~$<z<$~2.25~\AA\ with a slope
  $\varepsilon_{5\sigma}' = 2.3 \pm 0.1$~eV/\AA , and hence
 $\lambda \approx 0.20$~eV.

From the inelastic currents we  calculate desorption rates by solving
the Pauli master equation for the transitions among the various levels
of the oscillator\cite{GaPeLu92}.  We include
contributions from $I_1$, $I_2$, and $I_3$\cite{approx}, and  vibrational relaxations are described by a current independent
relaxation rate, $\gamma=1 \times
10^8$~s$^{-1}$\cite{ShWaAbTuLyAvWa95,GuLiMi95}. We assume that 
desorption  occurs  when the energy of the
H atom exceeds the desorption energy
$E_{\rm des}=3.36$~eV, corresponding to a truncated
harmonic potential well  with 13 levels.
The solid lines in Fig.~3(inset) show the calculated desorption rate as
function of tunnel current  for $V_{\rm b}=-5$~V and $-7$~V. The
agreement with the experimental data is excellent. The desorption rate
follows a powerlaw with exponent $\alpha\approx 6.5$. An  analysis
of the theoretical calculation shows that  the contribution from $I_2$
dominates the desorption rate, and therefore $\alpha \approx N/2$,
where $N=13$ is the number of levels in the truncated harmonic
potential well.  This contrasts with
the
low positive  bias case, $V_{\rm b} < 4$~V, where $I_1$
dominates and $\alpha \approx 13$\cite{ShWaAbTuLyAvWa95}. This
difference is due to the different lifetimes of the 5$\sigma$ and
6$\sigma^*$ resonances.  Since the 6$\sigma^*$ resonance is more
shortlived($\Delta=1$eV), the energy transfer in each
inelastic scattering event is smaler and $I_1$ dominates in this
case\cite{Stetal98}. Furthermore,  electrons with energy $<4$~eV  only sample the low
energy tail of the Si-H 6$\sigma^*$  resonance, and $\rho_1$ is
relatively broader than  $\rho_n$ for $n>1$, see Fig.~1.

We next calculate
$I_{\rm des}(V_{\rm b})$ and the results are shown in Fig.~3, and 
 we note that the minimum at $-7$~V is accurately
reproduced by
the theoretical calculation. This minimum does not
coincide with the resonance energy $\varepsilon_{5\sigma}$, but with the onset of
the field emission regime, as illustrated by  $I-V$
characteristics for this system at a constant field of 0.8~V/\AA\
(dashed line). In field 0.8~V/\AA\ surface states are bandbent by
$\sim 1$eV\cite{StQuGr97} and therefore
$\varepsilon_{5\sigma}\approx6$ eV. Thus, in the range $[-6.5{\rm
  V}$,$0 {\rm V}]$ the bias dependence of $I_{\rm des}$ is
mainly determined by the shape of the resonance wavefunction.  Below
$-6.5$V the bias dependence  is a result of competition
between   two different effects: polarization of the hole resonance
and increasing Fermi-level conduction. The electric field
polarizes the surface, displacing wave functions towards the
tip. Since the 
  $5\sigma$ orbital is more  polarized than bulk states, its overlap
  with the tip state increases relative to other states, and the fraction
  of electrons causing double vibrational excitations, 
  $f_2=I_2/I$, increases with electric field, as observed in the
  voltage range $[-7{\rm V}$,$-6.5 {\rm V}]$. On the other hand, the
 growing  tip-sample 
 distance with increasing magnitude of the  negative bias will
 increase the tunnel barrier of the  $5\sigma$ state 
 relative to states near the Fermi level,  thus  $f_2$
 decreases with tip-sample distance. This effect becomes 
dominant at the onset of the field emission regime,  where 
the conduction of states near the Fermi level 
is independent of the tip-sample distance, and 
$f_2$ starts to decrease when $V_{\rm b} <  -7$~V.

In conclusion, we have presented experimental   measurements of
the voltage and current dependent variation of the hydrogen desorption
rate from the Si(100)-H(2$\times1$) surface at negative bias
conditions. Based on a novel first principles theory
of inelastic scattering, we have shown that the desorption is caused
by  vibrational heating of the H atom due to  inelastic-scattering
with  the Si-H $5\sigma$ hole resonance.

This work was supported by  the Japanese Science and
Technology Agency and the  Danish Ministries of Industry and
Research in the framework of the
Center for Nanostructures(CNAST). The use of national computer
resources was supported by the Danish Research Councils.

\begin{figure}
\begin{center}
\leavevmode
\epsfxsize=85mm
\epsffile{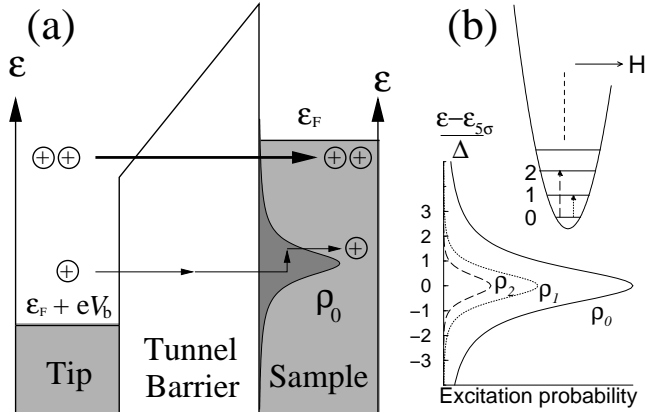}
\end{center}
\caption{(a) Inelastic tunneling of a hole into an adsorbate induced hole resonance with
  density of states, $\rho_0$. The higher barrier for 
  tunneling into the hole resonance compared to tunneling into
  Fermi-level states, means that only a fraction of the total tunnel
  current will pass through the hole resonance. (b) Schematic
  illustration of  relative energy dependent probabilities,
  $\rho_n(\varepsilon )$, for inelastic hole tunneling with energy
  transfer $n \hbar \omega_0$ to the adsorbate. } 
\end{figure}

\begin{figure}
\begin{center}
\leavevmode
\epsfxsize=65mm
\epsffile{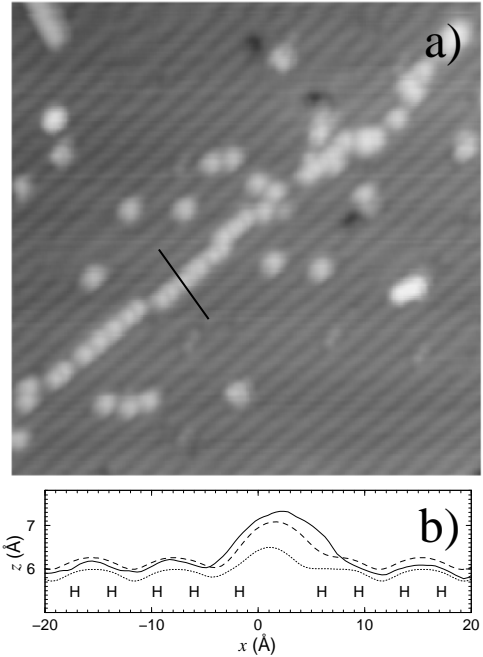}
\end{center}
\caption{(a) A  single line of desorbed H from the
Si(100)-H(2$\times$1) surface as a result of a
line scan at $-7$~V and 3.0~nA. 
 The picture is obtained at $-2$~V
and 0.2~nA, and  bright regions are due to  increased density of
states of H-free Si atoms.
(b) The experimental
corrugation(solid line) along the line in (a) compared with
the theoretical corrugation of a missing H defect including electric
field effects (dashed
line)  and without electric field effects (dotted line).
Results in the range $-7{\rm \AA } <x<9 {\rm \AA }$ are calculated using a
 c(4$\times$4)    cell  with one missing H atom and outside this range
 using  a (2$\times$1) cell. }
\end{figure}

\begin{figure}
\begin{center}
\leavevmode
\epsfxsize=85mm
\epsffile{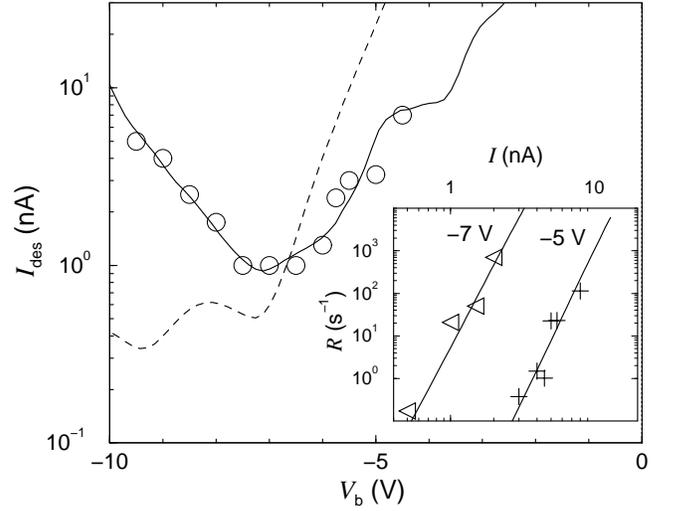}
\end{center}
\caption{ Current, $I_{\rm des}$,  as a function of sample bias,
  $V_{\rm b}$,  for constant
  desorption rate $R=4$~s$^{-1}$. Circles show experimental results, and
  the  solid line shows  the theoretical result.
 The dashed line shows
  the current  as a function of sample bias for constant electric
  field $E=0.8$~V/\AA\ in the tunnel gap. Inset shows $R(I)$  for
  $V_{b}=-7$~V(triangles) and $-5$~V(crosses), and 
  from least-squares fits of 
  $R\propto I^\alpha$ to the data we obtain $\alpha=5.7\pm0.7$~($-7$~V) and
  $6.3\pm1.3$~($-5$~V).    Lines show
  theoretical calculations   and have exponents  $\alpha=6.4$~($-7$~V)
  and  $7.1$~($-5$~V).} 
\end{figure}

\begin{figure}
\begin{center}
\leavevmode
\epsfxsize=85mm
\epsffile{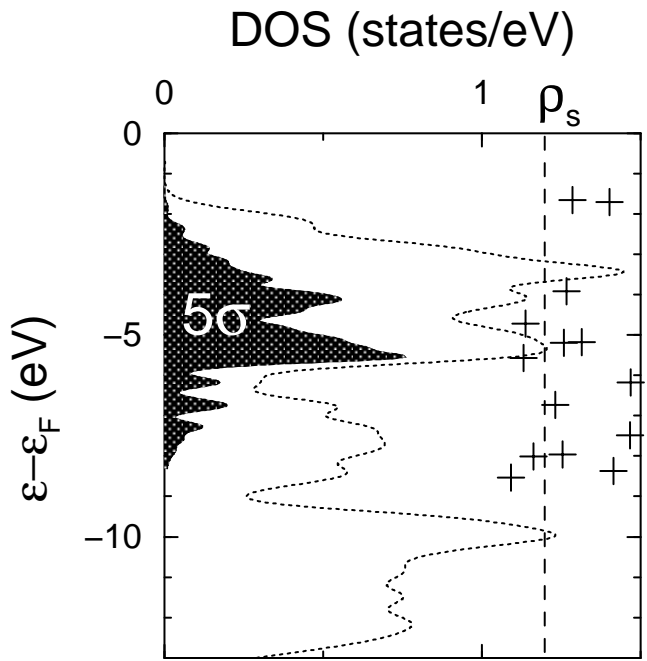}
\end{center}
\caption{a) The partial DOS, $n_{5\sigma}(\varepsilon)=\sum_\mu
  |\langle 5\sigma|\mu\rangle|^2\delta(\varepsilon-\varepsilon_\mu)$, of the
5$\sigma$ wave function  of a  Si-H molecule. The
dotted line shows the projection onto all Si-H molecular wave
functions. Crosses show values of $n_{5\sigma}(\varepsilon_\mu)/\langle 5\sigma|\mu\rangle|^2$.}
\end{figure}

\end{multicols}  

\begin{thebibliography}{10}
\bibitem{approx} 
   $I_n$ diverge
   for large $n$\cite{SaPePa94}, and the  divergence depends
   on $\lambda/\Delta$. In  the present calculation we use
  $n_{max}=3$, but obtain   similar results using   $n_{max}=2,3$ or $4$.
\end{thebibliography}
\end{document}